\documentclass[12pt]{article}

\usepackage{amsmath}         
\usepackage{amssymb}         
\usepackage{amsfonts}        
\usepackage{graphicx}        
\usepackage{physics}
\usepackage{tensor}
\usepackage{slashed}
\usepackage{cite}

\usepackage[
  pdftitle={},
  pdfauthor={},
    colorlinks=true,
    citecolor=black,
    linkcolor=black,
    urlcolor=blue,
    hypertexnames=false]{hyperref}

\usepackage[margin=2cm]{geometry} 

\usepackage{tikz}
\usetikzlibrary{decorations.markings}
\usetikzlibrary{automata, arrows}

\tikzset{->-/.style={decoration={
  markings,
  mark=at position #1 with {\arrow{>}}},postaction={decorate}}}

\newcommand{\beq}{\begin{equation}}
\newcommand{\eeq}{\end{equation}}

\newcommand{\qbar}{\overline q}
\newcommand{\Bbar}{\overline B}
\newcommand{\diag}{\text{diag}}

\newcommand{\arXiv}[2]{\href{http://arxiv.org/pdf/#1}{{\tt #2/#1}}}
\newcommand{\arXivold}[1]{\href{http://arxiv.org/pdf/#1}{{\tt #1}}}

\numberwithin{equation}{section}

\def\preprint{UCI-TR-2019-08}

\begin{document}

\begin{titlepage}

\def\thepage {}
\title{Deformations of the moduli space and\\ superpotential flows in 3D SUSY QCD}

\author{Yuri~Shirman\footnote{yshirman@uci.edu}~ and Michael~Waterbury\footnote{mwaterbu@uci.edu}}

\date{\footnotesize Department of Physics and Astronomy\\ University of California, Irvine, CA 92697}

\maketitle
\vspace*{-80mm}
\begin{flushright}
\preprint\\
\end{flushright}
\vspace*{65mm}

\begin{abstract}
\vspace*{.2in}
\noindent

We study the moduli space of three-dimensional $\mathcal{N}=2$ SQCD with $SU(N)$ gauge group and $F<N$ massless flavors.
In the case of an $SU(2)$ theory with a single massless flavor, we explicitly calculate the quantum constraint $YM=1$ and generalize the calculation to models with arbitrary $N$ and $F=N-1$ flavors.
In theories with $F<N-1$ flavors, we find that analogous constraints exist in locally defined coordinate charts of the moduli space.
The existence of such constraints allows us to show that the Coulomb branch superpotential generated by single monopole effects is equivalent to the superpotential generated by multi-monopole contributions on the mixed Higgs-Coulomb branch.
As a check for our result, we implement the local constraints as Lagrange multiplier terms in the superpotential and verify that deformations of a theory by a large holomorphic mass term for the matter fields results in a flow of the superpotential from the $F$-flavor model to the superpotential of an $(F-1)$-flavor model.
\end{abstract}

\end{titlepage}

\section{Introduction}
The study of non-perturbative dynamics of SUSY gauge theories in three and four dimensions lead to invaluable insights in understanding quantum field theories. An especially important role in non-perturbative dynamics is played by instantons in 4D  and monopoles\footnote{Topological instantons in 3D non-Abelian field theories are related to 4D monopoles and are often referred to as instanton-monopoles or simply monopoles.} in 3D. The exact non-perturbative superpotential in 3D $\mathcal{N}=2$ SUSY QCD was first calculated on the Coulomb branch of a pure SYM $SU(2)$ theory in \cite{Affleck:1982as}. An analogous instanton calculation in 4D $\mathcal{N}=1$ SUSY QCD with $SU(N)$ gauge group and $F=N-1$ flavors in a fundamental representation lead to a celebrated ADS superpotential \cite{Affleck:1983rr,Affleck:1983mk}. Further breakthroughs came in the 1990s with the discovery of dualities in 4D theories with $\mathcal{N}=2$ SUSY \cite{Seiberg:1994rs,Seiberg:1994aj} as well as in theories with  $\mathcal{N}=1$ SUSY \cite{Seiberg:1994bz,Seiberg:1994pq}. Understanding the interplay between instantons, monopoles, global symmetries, anomalies and weakly coupled limits allowed one to understand IR dynamics in a broad class of non-abelian gauge theories. Similar advances have been made in the study of theories on $\mathbf{R}^3\times \mathbf{S}^1$ and $\mathbf{R}^3$ \cite{Seiberg:1996nz,Intriligator:1996ex,deBoer:1997kr,Aharony:1997bx,Aharony:2013dha,Csaki:2014cwa,Csaki:2017mik,Csaki:2017cqm,Lee:1998bb,Poppitz:2013zqa,Dorey:1998kq,Dorey:1997ij,Aharony:1997gp}. It is often the case that a better understanding of the theory in 3D or 4D provides feedback to better understand the theory in 4D or 3D, respectively.

Depending on the number of flavors, SUSY QCD with four supercharges exhibits a rich range of phenomena in 3D and 4D, ranging from non-perturbative Affleck-Dine-Seiberg (ADS) superpotentials to quantum deformed moduli spaces to confinement without chiral symmmetry breaking and duality. The existence of quantum deformations was first discovered in a 4D $SU(N)$ theory with $F=N$ flavors. In this theory, the dynamical superpotential is not allowed by symmetries, since all gauge invariant moduli have charge $0$ under anomaly free R-symmetry. The moduli space of light fields can be described in terms of gauge invariant mesons, $M_{ij}$, and baryons, $B$ and $\Bbar$. The moduli space is subject to a classical constraint $\det M=B\Bbar$. It was shown by Seiberg \cite{Seiberg:1994aj} that non-perturbative dynamics deform the moduli space, and the constraint becomes $\det M-B\Bbar=\Lambda^{2N}$. This constraint can be most easily calculated in the $SU(2)$ theory with $2$ flavors by calculating a two-point chiral scalar correlation function on the Higgs branch in the constraint instanton background \cite{Finnell:1995dr} (see also \cite{Vainshtein:1986ns}). The low energy physics of the theory can be described in terms of the gauge invariant mesons and baryons subject to the quantum modified constraint which is enforced by a Lagrange multiplier term in the superpotential. This is consistent with the argument that the superpotential is not allowed by symmetries, since the equations of motion for the non-dynamical Lagrange multiplier force the superpotential to vanish. One could obtain the ADS superpotential for $F=N-1$ flavors by adding one mass term to the Lagrange multiplier superpotential and integrating out the massive flavor.  One could continue this integrating out procedure and obtain the superpotential for models with $F<N-1$ by adding mass terms for additional flavors. In the $F=N-1$ model, one could also obtain the ADS superpotential directly by calculating a two-point chiral fermion correlation function \cite{Affleck:1983rr,Affleck:1983mk,Cordes:1985um}. No direct calculation of the superpotential is possible in a theory on $\mathbf{R}^4$ when $F<N-1$. On the other hand, when the theory is compactified on a circle, multi-monopoles contribute to two-point fermion correlation functions and generate the pre-ADS superpotential \cite{Csaki:2017mik} of the 3D theory. Upon taking the large radius limit one recovers the ADS superpotential with an arbitrary number of flavors. This is one example when the study of 3D dynamics gives insight into dynamics of 4D theories.

In this paper, we will study quantum deformations of the classical moduli spaces in 3D SUSY QCD with $F<N$ flavors and investigate their role in the origin of the pre-ADS superpotentials as well as their role in the flow of superpotentials in the theory space as one adds holomorphic mass terms and decouples heavy flavors. It has long been known \cite{Aharony:1997bx} that the classical moduli space is deformed quantum mechanically in a 3D $SU(N)$ theory with $F=N-1$ flavors, taking the form $Y\!\det M=g^{2F}$, where $Y$ is a globally defined monopole modulus and $g^2$ is a 3-dimensional coupling. We will derive this constraint by following the approach of \cite{Finnell:1995dr} and calculating the two-point holomorphic scalar correlation function in an $SU(2)$ theory with one flavor. In fact, the physics behind such a modification is clearer in 3D. An $SU(2)$ theory with $4$ supercharges has a Coulomb branch along which the gauge group is broken to $U(1)$. On the Coulomb branch, the Higgs direction is lifted, and the squark vacuum expectation values (VEVs) are not allowed. Thus, the meson VEV, $M=q\qbar$, must vanish classically. Nevertheless, the holomorphic two-point squark correlation function receives nonvanishing contributions in a single monopole background. Indeed, the fundamental monopole of the $SU(2)$ theory has two gaugino and two doublet zero modes. Thus it generates a four-fermion vertex in the low energy effective theory. When this 't Hooft operator is combined with supersymmetric gauge couplings, one can construct a two loop diagram contributing to the two-point scalar correlation function. Such a diagram is naively UV divergent, but this divergence is cutoff by the finite size of the monopole. We perform a full calculation of this two-point correlation function in Sec. \ref{sec:F=N-1}. Then we generalize the result to $SU(N)$ theories with $F=N-1$ flavors and arbitrary $N$. In Sec. \ref{sec:F<N}, we observe that the classical moduli space is also deformed in theories with an arbitrary number of flavors, $F\le N-1$. When $F=N-1$ the deformation is global (constraining the global moduli, $Y$ and $M$), while in the case of $F<N-1$, the deformations exist in locally defined coordinate charts of the moduli space.
These local deformations lead to several important consequences. They guarantee the equivalence of the Coulomb branch superpotential discussed in \cite{Aharony:1997bx,Aharony:2013dha} and the multi-monopole generated superpotential on the mixed Higgs-Coulomb branch of the theory found in \cite{Csaki:2017mik}. Furthermore, the constraints ensure that the superpotential is valid in all coordinate charts on the moduli space. In Sec. \ref{sec:su3f1}, we present detailed analysis of the quantum moduli space of an $SU(3)$ model with $F=1$. In Sec. \ref{secsun1}, we extend our results to all $SU(N)$ models with $F=1$. Finally, in Sec. \ref{sec:sunf}, we generalize our discussions to arbitrary $F<N$ and show how the existence of such local deformations explains the superpotential flow between theories with different numbers of flavors as mass terms are added. We finish with a summary of our results in Sec. \ref{sec:conc}.

\section{Review of $3D$ $\mathcal N=2$ SUSY gauge theories} \label{sec:bkg}
In this section, we review basic properties of 3D SUSY QCD with four supercharges (${\mathcal N}=2$) (see, for example, \cite{Aharony:1997bx,deBoer:1997kr} for a more detailed introduction).
We restrict our attention to $SU(N)$ theories with $F<N$ massless flavors in the fundamental representation. The 3D action can be easily obtained by a dimensional reduction of the corresponding 4D theory:
\begin{equation}
    S = \int d^3x \left [\int d^4 \theta K(Q,\bar Q, V) + \int d^2\theta W(Q,\bar Q) + \frac{1}{g^2}\int d^2\theta \Tr(W_\alpha W^\alpha) + \text{H.c.} \right ]\,.
\end{equation}
We use supersymmetric normalization with an explicit factor $1/g^2$ in front of the gauge kinetic term. In this normalization, the vector multiplet has the same mass dimension as in 4D, since the gauge coupling, $g^2$, has mass dimension one in 3D. On the other hand, the chiral multiplet has mass dimension 1/2.
Expanding in component fields, the vector multiplet is given by
\begin{equation}
  V = -i\theta \bar\theta \sigma - \theta \gamma^i \bar\theta A_i + i \bar\theta^2 \theta \lambda - i \theta^2 \bar\theta \lambda^\dag + \frac{1}{2}\theta^2\bar\theta^2 D\,,
\end{equation}
where $\gamma^i=\{i \sigma^2,\sigma^1,\sigma^3\}$ and $\sigma$ is the real scalar field in the adjoint representation, and the chiral multiplet is given by
\begin{equation}
  Q = q + \psi\theta + \theta^2 F\,.
\end{equation}

The classical moduli space of the pure $\mathcal{N}=2$ SYM $SU(N)$ theory is described by the Coulomb branch parameterized by VEVs of the adjoint, $\ev{\sigma}=\diag(v_1,\ldots,v_{N})$, subject to the tracelessness condition, $\sum_i v_i=0$. At a generic point on the Coulomb branch, the unbroken gauge symmetry is $U(1)^{N-1}$. The theory on the Coulomb branch retains the Weyl symmetry of $SU(N)$ which interchanges the eigenvalues of $\sigma$. Without loss of generality we will restrict our attention to a positive Weyl chamber\footnote{The Weyl chamber is a wedge subspace of $\mathbb R^r$ ($r=rank(G)$) given by $\mathbb R^r/\mathcal W$, where $\mathcal W$ is the Weyl group \cite{Aharony:1997bx}.  The equivalence class from modding out the Weyl symmetry can be represented by a choice of $r$ positive, simple roots, $\{\alpha_i\}$, such that $\alpha_i \cdot \ev{\sigma} \geq 0$ or equivalently $v_i\geq v_{i+1}$. Sometimes this is referred to as the positive Weyl chamber \cite{Tong:2005un}.} defined by $v_i\geq v_{i+1}$. Quantum effects further divide the Weyl chamber into subwedges defined by the number of positive eigenvalues. We define the $k$th subwedge by requiring there to be exactly $k$ positive eigenvalues $(v_k>0>v_{k+1})$.
The subwedge boundaries lie at the points where one of the eigenvalues of $\sigma$ vanishes. We will call the boundary between the $k$th and $(k+1)$st subwedges the $k^{th}$ boundary. When $l=dim(ker(\ev{\sigma}))>1$, i.e. when several eigenvaues of $\sigma$ vanish simultaneously, $l-1$ subwedges become degenerate, and the symmetry breaking pattern is $SU(N)\longrightarrow U(l)\times U(1)^{N-l-1}$.

As first realized by Polyakov \cite{Polyakov:1976fu}, Abelian gauge theories without charged matter fields have a dual description in terms of compact scalar fields. The compact scalar fields, $\gamma$, obey the relation,
\begin{equation}
  \partial_i \gamma = \frac{\pi}{g^2}\epsilon_{ijk} F^{jk}\,,
\end{equation}
where $\gamma$ has a shift symmetry from its role as a Lagrange multiplier enforcing the Bianchi identity. In supersymmetric theories, this duality  provides Abelian vector superfields with a dual description in terms of chiral superfields, $\Phi$, with scalar components, $\phi = 4\pi \sigma/g^2 + i \gamma$.
The compactness of $\gamma$ ensures that the low energy theory depends on chiral superfields $Y=\exp(\Phi)$ charged under the global symmetry $U(1)_J$corresponding to the shift symmetry of $\gamma$.

The duality can be easily generalized to the Coulomb branch of non-Abelian gauge theories. In the case of an $SU(N)$ theory, the group is broken to a product of $N-1$ $U(1)$'s and each is dualized to a chiral superfield, obtaining a description in terms of $N-1$ chiral superfields, $Y_i$, defined as
\begin{equation}
Y_i=\exp\left[2\Tr(\phi T^i)\right]\,.
\end{equation}
The $(T^i)_{ab}=\frac{1}{2}(\delta_{a,i}-\delta_{a,i+1})\delta_{ab}$ are the generators of the corresponding non-orthogonal $U(1)$ subgroups of $SU(N)$.\footnote{We chose to describe the low energy $U(1)^{N-1}$ theory in terms of linearly independent but non-orthogonal $U(1)$ factors so that $Y_i$ are easily identified with fundamental monopoles of the positive Weyl chamber. One could also give a basis independent description of the Coulomb branch moduli in terms of the positive simple roots, $\{\alpha_i\}$, where $Y_i = \exp(\vec \alpha_i \cdot \phi)$.} After the duality transformation, the theory has no gauge symmetry. Instead, the moduli $Y_i$ are charged under topological global symmetries $U(1)_{J_i}$ associated with each abelian factor in the original gauge theory. The topological symmetry is broken by non-perturbative dynamics and is not a symmetry of the low energy physics.

On the Coulomb branch of the 3D theory, there exist monopole solutions charged under the corresponding $U(1)_{J_i}$ factors. All of the monopole and multimonopole solutions on the Coulomb branch can be constructed out of $N-1$ fundamental monopoles. In the positive Weyl chamber, the fundamental monopoles are the monopoles charged under one of the $U(1)_{J_i}$'s corresponding to the Abelian factor generated by $T^i$. The action of the $i$th fundamental monopole is given by
\begin{equation}
    S_{i,cl} = \frac{4\pi (v_i-v_{i+1})}{g^2}\,.
  \end{equation}
Comparing the classical monopole action with the VEVs of the Coulomb branch moduli $Y_i$, we note that the monopole weights are given by\footnote{Thus we will refer to $Y_i$'s as monopole moduli.} $1/Y_i$.

If we add $F$ massless flavors of chiral superfields in the fundamental representation of $SU(N)$, the theory possesses mixed Higgs-Coulomb branches of the moduli space (and a pure Higgs branch when $F \ge N-1$) in addition to the Coulomb branch. The mixed branch of the moduli space is not accessible from a generic point on the Coulomb branch -- the flat directions parameterized by squark VEVs are lifted by the D-term potential. Squark VEVs are only classically allowed when one or more $v_i$'s vanish.

First consider the case when only one adjoint VEV, say $v_k$, vanishes. The unbroken gauge symmetry is $U(1)^{N-1}$.
Of the $2NF$ chiral superfields, $2(N-1)F$ of them obtain large real masses, and the low energy effective theory is left with $2F$ massless chiral superfields.
These fields are charged under one linear combination of the unbroken $U(1)$'s,  and their VEVs must obey the D-flatness condition $\sum_f\left(|q^k_f|^2-|\qbar^k_f|^2\right)=0$. By flavor symmetry transformations, the squark VEVs can be rotated into a single flavor.
Alternatively, we could parameterize the vacua by the VEV of the meson superfield, $M$, which is classically defined as $M=Q\bar Q$ and has maximal rank one in this region of the moduli space.
Thus the space of physically inequivalent vacua on this branch is $N-1$ dimensional and can be parameterized by $N-2$ independent combinations of monopole moduli $Y_i$ and a single eigenvalue of $M$.

Now consider the case when several adjoint VEVs, say $l\le F$, vanish simultaneously. As discussed above, the unbroken gauge group in this region of the Coulomb branch is $U(l)\times U(1)^{N-l-1}$. The low energy physics contains $2lF$ massless chiral multiplets, and D-flatness conditions allow squark VEVS which further break the gauge group to $U(1)^{N-l-1}$. The meson matrix has rank $l$ and once again coordinates along the Higgs direction of this mixed branch can be parameterized by the $l$ eigenvalues of $M$. As before, the space of physically inequivalent vacua is $(N-1)$ dimensional.

It may also be useful to approach $U(1)^{N-l-1}$ low energy theory from a different direction on the classical moduli space. If we start at the origin of the classical moduli space, the entire $F^2$-dimensional Higgs branch is accessible, and one can turn on $l\le F$ meson eigenvalues breaking $SU(N)$ to $SU(N-l)$. At this point, an $(N-l-1)$-dimensional subspace of the Coulomb branch is accessible, and the gauge symmetry is further broken to $U(1)^{N-l-1}$. The space of physically inequivalent vacua remains $(N-1)$ dimensional. Of course, the superpotential of low energy theory in the $q\gg v$ limit should be the same as the superpotential in $q\ll v$ limit. Considering the theory in different VEV limits can be used as a tool to both to derive and verify our results.

The introduction of matter fields into the theory has one more important consequence: the $N-1$ fundamental monopole moduli are no longer globally defined throughout the Weyl chamber
because the quantum numbers of the moduli change as one crosses the boundary between different subwedges of the Weyl chamber. To understand this change of quantum numbers, we need to recall that quantum numbers of the monopole moduli depend on fermionic zero modes that exist in the background of the corresponding fundamental monopoles. Each fundamental monopole has two gaugino zero modes; however, only one fundamental monopole has matter fermion zero modes in any given subwedge of the Weyl chamber. For instance, in the $k$th subwedge (denoted by a superscript), $Y^{(k)}_k$ has one zero mode for each massless fundamental (or antifundamental) fermion, while $Y^{(k)}_i$ ($i\ne k$) has no matter fermion zero modes. The quantum numbers of mesons and fundamental monopoles in the $k$th subwedge are
\begin{equation}
 \begin{array}{| c | c | c | c | c | c |}
  \hline
          & U(1)_R & U(1)_B & U(1)_A & SU(F) & SU(F)   \\
  \hline
   Q      & 0 & 1  & 1 & \mathbf{F} & \mathbf{1} \\
   \bar Q & 0 & -1 & 1 & \mathbf{1} & \mathbf{\overline F} \\
   \hline
   M      & 0 & 0  & 2 & \mathbf{F} & \mathbf{\overline F} \\
   \hline
   Y^{(k)}_{k} & 2(F-1) & 0 & -2F & \mathbf 1 & \mathbf 1 \\
   Y^{(k)}_{i\ne k} & -2 & 0 & 0 & \mathbf 1 & \mathbf 1 \\ \hline
 \end{array}
\end{equation}
We can see that the quantum numbers of $Y^{(k)}_k$ and $Y^{(k)}_{k+1}$ monopoles in the $k$th subwedge are different from $Y^{(k+1)}_k$ and $Y^{(k+1)}_{k+1}$ in the $(k+1)$st subwedge despite the fact that both pairs of coordinates correspond to the same semiclassical solutions in each subwedge. One can define a two-monopole modulus which is continuous across the $k$th subwedge boundary,
\begin{equation}
    Y^{(k)}_{k,2}=Y^{(k)}_{k} Y^{(k)}_{k+1}=Y^{(k+1)}_{k} Y^{(k+1)}_{k+1}=Y^{(k+1)}_{k,2}\,.
\end{equation}
The introduction of the two-monopole modulus smooths out one combination of the two discontinous coordinates at each subwedge boundary. Specifically, the $Y^{(k)}_{k,2}$ modulus is still discontinuous at the $(k-1)$st subwedge boundary, but a different two-monopole modulus $Y^{(k)}_{k-1,2}$ is continuous at this boundary. It is possible to define a separate two-monopole modulus for each subwedge boundary that is continuous across that specific subwedge boundary.

One may hope to patch together the coordinate charts for each subwedge in this manner, but there are two technical issues which prevent such a construction. The first issue is the existence of a second modulus that is discontinuous at both subwedge boundaries. In other words, so far we have been able to define only one transition function for two discontinuous coordinates. A single continuous two-monopole modulus does not account for the two discontinuous monopole moduli.\footnote{One could define the global modulus, $Y=\prod_i Y_i$, which is continuous across all subwedge boundaries as has been done in previous studies \cite{Aharony:1997bx,Aharony:2013dha}. However, working only in terms of globally defined moduli does not allow one to investigate dynamics in the interior of the moduli space.} The other issue is that the two adjacent subwedges of the classical Coulomb branch do not overlap, so the transition functions can not be properly defined. As we will see, the quantum deformations of the classical moduli space solve these issues by smoothing out the Higgs-Coulomb interface at each of the subwedge boundaries. The extension of disjoint subwedges onto the intermediate Higgs-Coulomb branch allows these extended subwedges to overlap, while implementation of the quantum deformation as a Lagrange multiplier term in the superpotential provides the second transition function.

\section{$F=N-1$: Quantum Deformed Moduli Space} \label{sec:F=N-1}

In this section, we derive the 3D quantum constraint by calculating two-point holomorphic squark correlation function in $SU(2)$ theory with one flavor and generalizing the result to $SU(N)$ with $F=N-1$ flavors. As discussed in the previous section, the classical moduli space of the $SU(2)$ theory has two one-dimensional branches: a Higgs branch parameterized by a squark VEV $q=\bar q$ (or, in a gauge invariant language, by the meson $M\sim q\qbar$) and a Coulomb branch parameterized by the VEV of the adjoint scalar component of the gauge multiplet. Along the Coulomb branch, the gauge symmetry is broken to $U(1)$, and it is convenient to describe the physics in terms of the monopole modulus $Y$. Classical Higgs and Coulomb branches only intersect at the origin of the moduli space. Therefore, on the Coulomb branch, the holomorphic squark-antisquark correlation function must vanish clasically. However, as we explicitly show below, this correlation function obtains a nonvanishing contribution $\ev{M}=\ev{q\bar q}=g^2/Y$ in the monopole background on the Coulomb branch. The corresponding semiclassical calculation is weakly coupled and under control for sufficiently large $v$. Holomorphy guarantees that this result remains valid everywhere on the Coulomb branch, implying a well known 3D quantum constraint $YM=g^2$.

\begin{figure}[h]
\begin{center}
    \begin{tikzpicture}[decoration={markings}]
        \begin{scope}
            \node [shape=circle,draw=black] (Y) at (0,0) {$Y$};
            \node [shape=circle,draw=black,fill=black,inner sep=0pt,minimum size=3pt] (y1) at (-1.5,0) {};
            \node [shape=circle,draw=black,fill=black,inner sep=0pt,minimum size=3pt] (y2) at (1.5,0) {};
            \node (x1) at (-3,0) {$q(x)$};
            \node (x2) at (3,0) {$\bar q(x)$};
        \end{scope}

        \begin{scope}[>=stealth',shorten >=0,node distance=2.8cm]
            \path [draw,->-=0.5] (Y) to [bend right=80] node[above] {$\lambda_0^{[1]}$} (y1);
            \path [draw,->-=0.5] (Y) to [bend left=80] node[below] {$\psi_0$} (y1);
            \path [draw,->-=0.5] (Y) to [bend left=80] node[above] {$\lambda_0^{[2]}$} (y2);
            \path [draw,->-=0.5] (Y) to [bend right=80] node[below] {$\bar \psi_0$} (y2);
            \path [draw,dashed] (y1) edge (x1);
            \path [draw,dashed] (y2) edge (x2);
        \end{scope}
    \end{tikzpicture}
\end{center}
\caption{Diagram illustrating the monopole contribution to squark correlation functions}
\label{singlepole}
\end{figure}
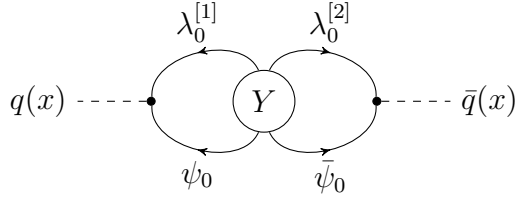

Our calculation is similar to 4D calculations of quantum constraints in \cite{Finnell:1995dr}. The instanton monopole of the $SU(2)$ theory with one flavor has two gaugino and two fundamental zero modes and contributes to chiral four-fermion correlation function. This correlation function can be converted to holomorphic two-point squark correlation function by the insertion of two supersymmetric gauge couplings. The resulting contribution can be visualized in Fig. \ref{singlepole}.

In the language of the path integral, we must evaluate
\begin{equation}
    \expval{\bar q^i(x) q_i(x)} = \int [\mathcal{D}\phi]_{qu} \left(\bar q^i(x) q_i(x)\right)e^{-S_{cl.} - S[\phi_{qu}]}\,, \label{pathint}
\end{equation}
where $[\phi_{qu}]$ is shorthand for all quantum field fluctuations around the monopole background. We present the details of the calculation in the Appendix. The resulting correlation function is
\begin{equation}
\ev{\bar q^i(x) q_i(x)} = \frac{v^2}{g^4}e^{-\frac{4\pi v}{g^2}} g^2 I\,,
\end{equation}
where $I$ is a positive definite integral. At first sight, this is nonholomorphic, but as explained in \cite{Dorey:1998kq,Poppitz:2012sw}, the nonholomorphic prefactor, $v^2/g^4$, can be absorbed into redefinition of the K\"ahler potential. The required field redefinition leads to a finite renormalization of the gauge coupling
\begin{equation}
 \frac{1}{g^2}\longrightarrow \frac{1}{g^2}-\frac{2}{v}\,.
\end{equation}
In terms of the rescaled modulus $Y$, the two-point scalar correlation function becomes
\begin{equation}
\label{squarkcorr:ans}
 \ev{M}=\frac{g^2}{Y}\,.
\end{equation}

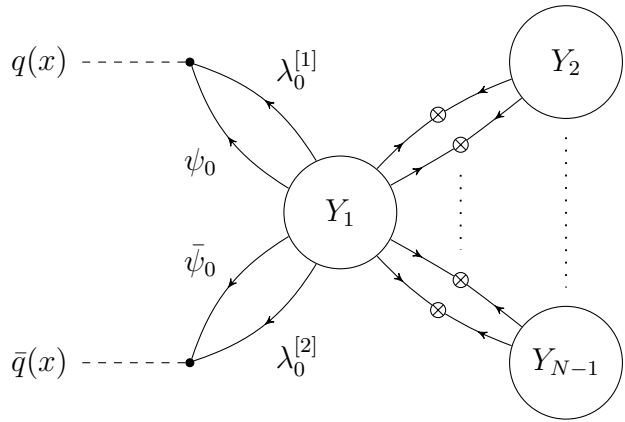
\begin{figure}[h]
\begin{center}
    \begin{tikzpicture}[decoration={markings},cross/.style={path picture={
  \draw[black]
(path picture bounding box.south east) -- (path picture bounding box.north west) (path picture bounding box.south west) -- (path picture bounding box.north east);
}}]
        \begin{scope}
            \node [shape=circle,draw=black,minimum size=1.5 cm] (Y1) at (0,0) {$Y_1$};
            \node [shape=circle,draw=black,minimum size=1.5 cm] (Y2) at (3,2) {$Y_2$};
            \node [shape=circle,draw=black,minimum size=1.5 cm] (YN) at (3,-2) {$Y_{N-1}$};
            \node [shape=circle,draw=black,fill=black,inner sep=0pt,minimum size=3pt] (y1) at (-2,2) {};
            \node [shape=circle,draw=black,fill=black,inner sep=0pt,minimum size=3pt] (y2) at (-2,-2) {};
            \node (x1) at (-4,2) {$q(x)$};
            \node (x2) at (-4,-2) {$\bar q(x)$};
            \node [draw,circle,cross,inner sep=0pt,minimum width=0.2 cm] (z11) at (1.3,1.3) {};
            \node [draw,circle,cross,inner sep=0pt,minimum width=0.2 cm] (z12) at (1.6,0.9) {};
            \node [draw,circle,cross,inner sep=0pt,minimum width=0.2 cm] (z21) at (1.3,-1.3) {};
            \node [draw,circle,cross,inner sep=0pt,minimum width=0.2 cm] (z22) at (1.6,-0.9) {};
        \end{scope}

        \begin{scope}[>=stealth',shorten >=0,node distance=2.8cm]
            \path [draw,->-=0.5] (Y1) to [bend right=20] node[above right] {$\lambda_0^{[1]}$} (y1);
            \path [draw,->-=0.5] (Y1) to [bend left=20] node[below left] {$\psi_0$} (y1);
            \path [draw,->-=0.5] (Y1) to [bend left=20] node[below right] {$\lambda_0^{[2]}$} (y2);
            \path [draw,->-=0.5] (Y1) to [bend right=20] node[above left] {$\bar \psi_0$} (y2);
            \path [draw,->-=0.5] (Y1) to [bend left=5] (z11);
            \path [draw,->-=0.5] (Y1) to [bend right=1] (z12);
            \path [draw,->-=0.5] (Y2) to [bend right=5] (z11);
            \path [draw,->-=0.5] (Y2) to [bend left=1] (z12);
            \path [draw,->-=0.5] (Y1) to [bend right=5] (z21);
            \path [draw,->-=0.5] (Y1) to [bend left=1] (z22);
            \path [draw,->-=0.5] (YN) to [bend left=5] (z21);
            \path [draw,->-=0.5] (YN) to [bend right=1] (z22);
            \path [draw,dashed] (y1) edge (x1);
            \path [draw,dashed] (y2) edge (x2);
            \path [draw,loosely dotted, line width=0.25 mm] (3,1) edge (3,-1);
            \path [draw,loosely dotted, line width=0.25 mm] (1.6,0.5) edge (1.6,-0.5);
        \end{scope}
    \end{tikzpicture}
\end{center}
\caption{Diagram illustrating the multimonopole contribution to a $2F$ squark correlation function}
\label{multipole}
\end{figure}

The generalization to $SU(N)$ theories with $F=N-1$ flavors is reasonably straightforward.
Consider the theory on the mixed Higgs-Coulomb branch where the rank of the meson $M$ is $N-2$ and the low energy physics is described by an $F=1$ $SU(2)$ theory. The calculation of the two-point holomorphic scalar correlation function is illustrated in Fig. \ref{multipole} where crosses represent VEV insertions. As before, the coupling constant of the low energy theory is renormalized due to the contributions of nonzero modes and is shifted by terms of the form $1/v$.
The fundamental monopole of the low energy $SU(2)$ can be written in terms of the fundamental monopoles and mesons of the high energy theory $Y_L=\prod_{i} Y_i\det^\prime M/g^{2(F-1)}$, where the prime denotes the determinant over $N-2$ flavors with nonvanishing VEVs.
The scalar correlation function for the remaining massless flavor of the low energy $SU(2)$ follows from our earlier calculation and gives the quantum constraint $Y\det M=g^{2F}$. One could also derive the constraint by calculating the $2(N-1)$-point scalar correlation function at a generic point on the Coulomb branch in the background of $N-1$ fundamental monopoles, but this calculation would be difficult in practice.

The existence of quantum deformations of the classical moduli space for an arbitrary number of flavors implies that the rank of the meson superfield is maximal ($rank(M)=N-1$). We will later see that $M$ will have maximal rank ($rank(M)=F$) for any number of flavors. This means that, quantum mechanically, the gauge symmetry is always maximally broken and some fundamental monopoles can not contribute to the superpotential despite the fact that, at a generic point on a pure Coulomb branch, these monopoles have exactly the two-fermion zero modes necessary to generate two-fermion correlation function and the corresponding superpotential terms.

\section{$F<N-1$: Quantum Constraints as Transition Functions} \label{sec:F<N}
As discussed in Sec. \ref{sec:bkg}, the space of physically inequivalent classical vacua consists of several distinct $(N-1)$-dimensional branches. One might expect that the moduli space becomes a smooth, locally connected manifold in the quantum theory.
We could attempt to describe such a manifold in terms of globally defined moduli $Y$ and $M$. However, $M$ has maximal rank $F$, and there are an insufficient number of globally defined moduli ($F+1$) to parameterize the entire $(N-1)$-dimensional moduli space when $F<N-1$.
The fundamental monopoles can not serve as additional coordinates on the moduli space, since they are discontinous at subwedge boundaries. To resolve the problem, one would need to introduce new composite coordinates valid in two or more subwedges and transition functions between the composite coordinates that are valid on overlapping sets of subwedges. For example, one could use the two-monopole modulus $Y_{k,2}^{(k)}$ discussed earlier. However, this is not sufficient, because there are two discontinous coordinates at each subwedge boundary. One of these coordinates can be replaced by the two-monopole modulus, but the existence of a second discontinous coordinate will prevent us from patching together disjoint subwedges.
As we will show below, quantum effects deform the classical moduli space even in theories with $F<N-1$, but such deformations are local (i.e. they are only valid in specific subwedges of the classical moduli space). Moreover, these deformations smooth out the interface between the subwedges and make the mixed Higgs-Coulomb branch accessible from either adjacent subwedge.
These quantum deformations also provide necessary transition functions to cover the entire quantum moduli space with overlapping coordinate charts. The quantum deformed moduli space is further lifted by monopoles, and the exact superpotential of the theory can be written down in terms of the appropriate coordinates in all coordinate patches.

In Sec. \ref{sec:su3f1}, we will show how this plays out in the case of an $SU(3)$ theory with $F=1$. While the $SU(3)$ example is illuminating, it is not sufficiently general. In this case, there are two globally defined moduli $Y=Y_1Y_2$ and $M$ which can describe dynamics everywhere on the moduli space. In Sec. \ref{sec:F=1}, we extend these results to all $SU(N)$ theories with $F=1$. We show that once again quantum effects deform the classical moduli space, relating monopole and meson moduli at each boundary. This deformation allows us to cover the moduli space by a set of overlapping coordinate charts, with each patch covering two neighboring subwedges. We will demonstrate that calculations of the pre-ADS superpotential generated by single monopole contributions in any subwedge of the Coulomb branch lead to the same result, and this is the same superpotential that can be found by considering monopole and multimonopole contributions on mixed Higgs-Coulomb branches emanating from the boundaries between subwedges.
In Sec. \ref{sec:sunf}, we further generalize the results to $SU(N)$ models with $F<N-1$ flavors.  Here the quantum deformation relates the mesons to $F$-monopole composite operators
\begin{equation}
Y_{k,F}\det M =\left(\prod_{i=k}^{k+F-1}Y^{(k)}_i\right)\det M=g^{2F}\,.
\end{equation}
 This local deformation allows us to introduce the $(F+1)$-monopole modulus $Y_{k,F+1}=\prod_{i=k}^{k+F+1}Y^{(k)}_i$ which is continuous across the intermediary boundaries $Y^{(k)}_{k,F+1}=Y^{(k+1)}_{k,F+1}=\ldots=Y^{(k+F)}_{k,F+1}$. We can then cover the moduli space by a set of overlapping coordinate charts with well defined transition functions and show that superpotentials calculated on all of the quantum-mechanically accessible branches of the moduli space are equivalent.

\subsection{$SU(3)$ theory with $F=1$}
\label{sec:su3f1}

Consider a one flavor $SU(3)$ model on the Coulomb branch in the positive Weyl chamber. Classically, the Weyl chamber is split into two subwedges depending on the sign of $v_2$, while at $v_2=0$, the Higgs branch is accessible. In the $v_2<0$ subwedge, it is convenient to parameterize the Coulomb branch coordinate by monopole moduli, $Y_1$ and $Y_2$, which in a semiclassical regime are approximated by $Y_1\sim\exp\left[4\pi(v_1-v_2)/g^2\right]$ and $Y_2\sim \exp\left[4\pi(v_2-v_3)/g^2\right]$.
In the $v_2>0$ subwedge, we need to choose different Coulomb branch moduli $Y_1^\prime$ and $Y_2^\prime$. As explained earlier, despite similar behavior in the semiclassical regime, the quantum numbers of $Y_1^\prime$ and $Y_2^\prime$ differ from those of $Y_1$ and $Y_2$ respectively and do not represent the same degrees of freedom in the quantum theory.

Consider the first subwedge of the Weyl chamber defined by $v_1>0>v_2>v_3$. Here the first fundamental monopole $Y_1$ has four (two gaugino and two matter) fermion zero modes, while the second fundamental monopole $Y_2$ has two gaugino zero modes. Only the second fundamental monopole contributes to the superpotential, and we find $W = Y_2^{-1}$. In addition, we can calculate two-point scalar correlation function in $Y_1$ background. Since the fields of the $Y_1$ monopole (including the fermion zero modes) can be embedded in an $SU(2)$ subgroup of $SU(3)$, the calculation is nearly identical to the one we performed in the previous section. There are two new features that must be taken into account. First, the contributions of nonzero modes of the matter doublets are modified since, from the point of view of the $Y_1$ monopole, the matter fermions have real mass $(v_1+v_2)/2$. Second, the gaugino contains components that transform as doublets in the $Y_1$ monopole background. These components of the gaugino do not have zero modes, but their nonzero modes contribute to the path integral just like matter doublets with a real mass $3(v_1+v_2)/2$ would. Similar to the nonholomorphic prefactor in the $SU(2)$ theory, these effects can be understood as finite renormalization of the $U(1)$ gauge coupling of the low energy theory, and we find $Y_1M\sim g^2$. This result can be enforced in the first subwedge of the Weyl chamber through a Lagrange multiplier term in the superpotential,
\begin{equation}
\label{eq:su3f1one}
 W=\frac{g^4}{Y_2}+\lambda_1 (Y_1M-g^2)\,.
\end{equation}
In the $v_1>v_2>0>v_3$ sub-wedge, we similarly find
\begin{equation}
\label{eq:su3f1two}
 W=\frac{g^4}{Y_1^\prime}+\lambda_2 (Y_2^\prime M-g^2)\,.
\end{equation}
To verify that these two expressions are consistent with each other we must compare them at a jumping point where $v_2=0$. Integrating out the Lagrange multiplier terms, both forms of the superpotential lead to the same result
\begin{equation}
\label{eq:su3f1}
 W=\frac{g^6}{Y_1Y_2M}=\frac{g^6}{Y_1^\prime Y_2^\prime M}\,,
\end{equation}
where the composite two-monopole modulus, $Y=Y_1Y_2=Y^\prime_1Y^\prime_2$, is continous across the boundary between the two subwedges as explained in Sec. \ref{sec:bkg}.
It is tempting to interpret the superpotential (\ref{eq:su3f1}) as a two-monopole contribution to the superpotential, and indeed it agrees with the results of the two monopole superpotential calculation on the mixed Higgs-Coulomb branch of the theory \cite{Csaki:2017mik}. We conclude that $Y$ and $M$ are valid in both coordinate charts of the $F=1$ $SU(3)$ theory and are related to moduli of the two coordinate charts by
$\{Y_1=g^2/M, Y_2=YM/g^2\}$ and $\{Y^\prime_1=YM/g^2,Y^\prime_2=g^2/M\}$.

Note that the above procedure is precisely that described at the beginning of this section. The quantum constraints enforced by $\lambda_1$ and $\lambda_2$ provided a transition functions from the two sets of Coulomb branch coordinates to the mixed Higgs-Coulomb branch coordinates, while the continuity of the two-monopole modulus ensures that the two Coulomb branch coordinate charts overlap on the Higgs branch. Together, these effects guarantee agreement between all three expressions for the superpotential.

Let us consider some standard checks of ADS and pre-ADS superpotentials. Specifically, we can study the theory on a Higgs branch where the low energy physics is described by a pure SYM $SU(2)$ as well as deform the theory by a large holomorphic mass term, $m$, so that the low energy description is given in terms of a pure SYM $SU(3)$ theory. In the former case, the low energy superpotential is given by $1/Y_{L}$, and by comparing with (\ref{eq:su3f1}), we conclude that the matching of high and low energy theories requires a rescaling of chiral superfields to absorb $M$ into the definition of $Y_L=Y_1Y_2(M/g^2)$. Similarly to nonholomorphic rescaling discussed  in \cite{Poppitz:2012sw}, this field redefinition affects the matching relation between the coupling constants of the high and low energy theories and should reproduce the finite renormalization of the low energy $U(1)$ coupling constant.
When the theory is deformed by a mass term, the low energy superpotential must be
\begin{equation}
\label{eq:su3sym}
 W=\frac{g_L^4}{Y_{1L}}+\frac{g_L^4}{Y_{2L}}\,.
\end{equation}
We can obtain this superpotential by starting either with (\ref{eq:su3f1one}) or (\ref{eq:su3f1two}) and adding a mass term. For example, in the $v_2<0$ subwedge of the Weyl chamber, the superpotential is
\begin{equation}
\label{eq:su3f1mass}
 W=\frac{g^4}{Y_2}+\lambda_1(Y_1M-g^2)+m M\,.
\end{equation}
Integrating out both the Lagrange multiplier and the meson superfield, we find the low energy superpotential,
\begin{equation}
 W=\frac{g^2 m}{Y_1}+\frac{g^4}{Y_2}\,,
\end{equation}
which agrees with (\ref{eq:su3sym}) if  we identify\footnote{Here we neglect finite non-holomorphic shifts in the coupling discussed earlier.} $g_L^2=g^2$, $Y_{1L}=Y_1g^2/ m$ and $Y_{2L}=Y_2$. Once again, the rescaling required to absorb the mass into the $Y_L$ monopole of the low energy theory determines the coupling constant matching and correctly reproduces the renormalization of the $U(1)_1$ coupling constant. We stress that the local deformation of the moduli space, implemented through the Lagrange multiplier term in (\ref{eq:su3f1mass}), plays an essential role in reproducing the superpotential of the SYM low energy theory when the matter fields are decoupled by taking the superpotential mass term, $m$, to infinity.

\subsection{$SU(N)$ with $F=1$} \label{sec:F=1}
\label{secsun1}
The generalization to $SU(N)$ theories with an arbitrary number of colors and one massless flavor is straightforward. We will denote monopole moduli  in the $k$th subwedge by $Y_i^{(k)}$, $i=1,\ldots,N-1$. With the exception of $Y^{(k)}_k$, all the fundamental monopoles in this subwedge have two gaugino zero modes and no matter zero modes. Thus they contribute to the superpotential. Calculating the two point scalar correlation function, we find a local constraint applicable to the $k$th subwedge, $Y_k^{(k)}M=g^2$.  Thus within this subwedge of the Coulomb branch, the physics is described by the superpotential
\begin{equation}
\label{eq:sunf1}
 W=\sum_{i\ne k}\frac{g^4}{Y_i^{(k)}}+\lambda_k(Y_k^{(k)}M-g^2)\,.
\end{equation}
Although this superpotential is calculated in the $k$th subwedge, it can be extended into the $(k+1)$st (or $(k-1)$st) subwedges by using the constraint as a transition function and replacing $Y_{k+1}^{(k)}$ (or $Y_{k-1}^{(k)}$) by the composite two-monopole modulus that is continuous in the appropriate regions.
Let us explicitly carry this out for the $(k+1)$st subwedge. Integrating out the Lagrange multiplier, the superpotential can be written as
\begin{equation}
 W=\sum_{i\ne k,k+1}\frac{g^4}{Y_i^{(k)}}+\frac{g^6}{Y^{(k)}_kY^{(k)}_{k+1}M}\,.
\end{equation}
The last term can be interpreted as arising from the two monopole contribution considered in \cite{Csaki:2017mik}. In this form, the superpotential is valid both in the $k$th and $(k+1)$st subwedges due to their overlap at the mixed Higgs-Coulomb boundary. However, in the $(k+1)$st subwedge, the same superpotential can be written in two more forms. First, it can be written in terms of the monopole moduli of the $(k+1)$st subwedge, $Y_i^{(k+1)}$, and the local constraint, $\lambda_{k+1}\left(Y^{(k+1)}_{k+1}M-g^2\right)$, valid in this subwedge:
\begin{equation}
  W=\sum_{i\ne k+1} \frac{g^4}{Y_i^{(k+1)}} + \lambda_{k+1}(Y_{k+1}^{(k+1)}M-g^2)\,.
\end{equation}
Second, it can be written in terms of the composite monopole moduli $Y^{(k+1)}_{k+1}Y^{(k+1)}_{k+2}M/g^2$. Recall that this term in the superpotential can be interpreted as a two monopole contribution generated on the mixed Higgs-Coulomb branch accessible from the boundary between the $(k+1)$st or $(k+2)$nd subwedges. This procedure can be used to recursively generate the sets of coordinate charts and transition functions required to cover the entire quantum moduli space of the theory and to define it as a smooth, locally connected manifold. Moreover, the calculations on all accessible branches of the moduli space lead to the same results. It is easy to see that, just like in the case of the $SU(3)$ theory, the deformation of the theory by the mass term correctly leads to the low energy physics described by a pure $\mathcal{N}=2$ SYM $SU(N)$ theory.

\subsection{$SU(N)$ with $F<N-1$}
\label{sec:sunf}

We conclude our study of the pre-ADS superpotentials and quantum deformations of the moduli space by considering a general case of an $SU(N)$ theory with $F$ massless flavors. We will consider the first sub-wedge of the Weyl chamber, $v_1>0\geq v_2 \geq \ldots \geq v_N$. By calculating $2F$ scalar correlation function in the $F$-monopole background, one finds there exists a local constraint given by $\left(\prod_{i=1}^F Y_i\right)\det M=g^{2F}$.\footnote{Similar to the $SU(N)$ with $F=N-1$ case described in Sec. \ref{sec:F=N-1}, the constraint enforces $rank(M)=F$ and prohibits the superpotential of $F$ individual fundamental monopoles.} This is easiest to see by performing a calculation on the mixed Higgs-Coulomb branch where the rank $F-1$ meson VEV is allowed.
This is the region where $v_i=0$ for $i=2,\ldots,F-1$.
In the presence of VEVs, the multimonopole, $Y_{1,F}^{(1)}=\prod_{i=1}^F Y_i^{(1)}$, will collapse into a single fundamental monopole of the low energy $SU(N-F+1)$ theory, $Y_{1L}=Y_{1,F}^{(1)}\det^\prime M/g^{2(F-1)}$, where $\det^\prime M$ denotes determinant taken over $F-1$ flavors with large VEV. As discussed earlier, the rescaling used in the definition of $Y_{1L}$ shifts the coupling of the low energy theory. We can calculate the two-scalar correlation function for the remaining squark flavor in the low energy effective theory and find
\begin{equation}
 \ev{M_{FF}}=\frac{g^2_L}{Y_{1L}}=\frac{g^{2F}}{Y_{1,F}^{(1)} \det^\prime M}\,.
\end{equation}
One can then write the full nonperturbative superpotential in the form,
\begin{equation}
\label{eq:sunfone}
 W=\sum_{i=F+1}^{N-1}\frac{g^4}{Y_i^{(1)}}+\lambda_1\left(Y_{1,F}^{(1)}\det M-g^{2F}\right)\,.
\end{equation}
As expected, integrating out the Lagrange multiplier term, we find the multimonopole generated superpotential found in \cite{Csaki:2017mik},
\begin{equation}
 W=\sum_{i=F+2}^{N-1}\frac{g^4}{Y_i^{(1)}}+\frac{g^{2F+4}}{Y_{1,F+1}^{(1)}\det M}\,,
\end{equation}
where $Y_{1,F+1}^{(1)}=\prod_{i=1}^{F+1}Y_i^{(1)}$.

Alternatively, we can consider the second subwedge, $v_1>v_2>0\geq\ldots\geq v_N$ where the rank of $M$ is $F-1$. Here we find the superpotential
\begin{equation}
\label{eq:sunftwo}
 W=\frac{1}{Y_1^{(2)}}+\sum_{i=F+2}^{N-1}\frac{g^4}{Y_i^{(2)}}+\lambda_2\left(Y_{2,F}^{(2)}\det M-g^{2F}\right)\,.
\end{equation}
The valid coordinate patches for (\ref{eq:sunfone}) and (\ref{eq:sunftwo}) overlap on the mixed Higgs-Coulomb branch where both superpotentials are
\begin{equation}
  W=\sum_{i=F+2}^{N-1}\frac{g^4}{Y_i^{(1/2)}}+\frac{g^{2F+4}}{Y_{1,F+1}^{(1/2)}\det M}\,.
\end{equation}
With the help of the constraints, we can construct transition functions that allow us to cover the full moduli space with coordinate charts and verify that superpotentials calculated in any of these charts are equivalent.

Finally, we deform the theory by adding the mass term $mM_{FF}$ to the last flavor. Integrating out the heavy flavor we find the superpotential of low energy $SU(N)$ theory with $F-1$ flavors
\begin{equation}
 W=\sum_{i=F+1}^{N-1}\frac{g^4}{Y_i^{(1)}}+\frac{m g^{2F}}{Y_{1,F}^{(1)}\det^\prime M}\,.
\end{equation}
This is precisely the superpotential of the $F-1$ flavor theory calculated in \cite{Csaki:2017mik}. In addition, we need to compliment this superpotential by a new local constraint $Y_{1,F-1}^{(1)} \det^\prime M=g^{2(F-1)}$. This superpotential can then be extended to other regions of the $F-1$ flavor theory moduli space or reduced to the superpotential of the $F-2$ flavor theory by adding another large mass term.

\section{Conclusions}
\label{sec:conc}
In this paper, we explicitly calculated quantum constraint $YM=g^2$ in the 3D $SU(2)$ theory with one massless flavor and showed how to generalize the calculation to an $F=N-1$ theory with an arbitrary $N$. We also showed that a local version of such a constraint exists in 3D $SU(N)$ theories with an arbitrary number of flavors, $F<N-1$ flavors. The existence of local constraints allowed us to construct a set of coordinate charts that cover the entire moduli space and show that the superpotential calculations in different charts are equivalent. Additionally, the existence of local constraints ensures the agreement between the superpotentials generated by fundamental monopoles on the pure Coulomb branch of an $SU(N)$ theory \cite{Aharony:1997bx} with the superpotentials arising from fundamental monopoles and multimonopole contributions on the mixed Higgs-Coulomb branch of the theory \cite{Csaki:2017mik}. The validity of the superpotential throughout the entire moduli space implies that the physics is fully described by a single Coulomb branch of the low energy pure SYM $SU(N-F)$ theory coupled to dilaton-like moduli, $M$. We also showed that constraints play an essential role in the flow of the superpotential between theories with different numbers of flavors. When a superpotential mass term for one flavor is added to the theory and the heavy flavor is decoupled, the local constraint guarantees that the low energy superpotential reproduces the one expected in a theory with $F-1$ flavors. To continue the flow in flavor space as additional mass terms are added, one must include new local constraints that are generated whenever new mass terms are added to the superpotential. We expect that our analysis of $SU(N)$ gauge theories by considering deformations of the classical moduli space will be useful in understanding gauge theories with more general gauge groups and matter content.

\section*{Acknowldegements}
This work is supported in part by NSF grant PHY-1620638.

\appendix
\section{Squark Correlation Function Calculation}

In this Appendix, we discuss the evaluation of the path integral for an $SU(2)$ theory with one flavor which leads to \eqref{squarkcorr:ans}. We begin by deriving the instanton monopole integration measure, collective coordinates, one-loop determinants and all. Then we derive the fermionic zero mode functions and evaluate the integral. As a preparation for this discussion, let us recall properties of the single monopole configuration corresponding to the first nontrivial solution of the classical equations of motion \cite{Polyakov:1974ek,tHooft:1974kcl},
\begin{equation}
\begin{gathered}
    A^a_i(r) = \epsilon^{aij} \frac{n^j}{r}F(vr)\,, \quad \sigma^a(r) = v n^a H(vr)\,, \\
    F(\rho) = 1 - \frac{\rho}{\sinh{\rho}}\,, \quad H(\rho) = \frac{\cosh \rho}{\sinh \rho} - \frac{1}{\rho}\,.
\end{gathered}
\end{equation}
Such solutions satisfy the lower bound of the Bogomol'nyi bound and are exact since the adjoint scalar has no classical potential. There remain quantum fluctuations of the fields in this classical background. Some of these fluctuations do not have a corresponding change in the action. These are the zero modes of the instanton monopole. Index theorems guarantee a certain number of zero modes.

Specifically, the single monopole has four bosonic zero modes: three for the position of the monopole and one for a leftover $U(1)$ transformation. If we add a fermion $\psi$ to the theory, it also acquires zero modes satisfying
\begin{equation}
    (i \mathcal D_\mu \bar\sigma^\mu)\psi = ( i \partial_i \sigma^i + A^a_i\sigma^i T^a + \sigma^a T^a)\psi = 0\,,
\end{equation}
where the number of zero mode solutions and $T^a$ depend on the representation of $\psi$. We normalize our generators such that $\Tr(T^aT^b)=\delta^{ab}/2$ in the fundamental representation. Fermions in the adjoint representation have two zero modes, and fermions in the fundamental representation have a single zero mode.

\subsection{Monopole-Instanton Measure}
As discussed in Sec. \ref{sec:F=N-1}, the monopole contribution to the squark correlation function is \eqref{pathint}. In this formula, all fields have been expanded around their classical solution, $\phi = \phi_{cl.} + \phi_{qu.}$. The quantum fluctuations come in two types: nonzero modes and zero modes.
To leading order, the nonzero mode fluctuations are gaussian and their evaluation reduces to determinants of $\Delta_- = \slashed{\mathcal D}_{cl}\bar{\slashed{\mathcal D}}_{cl}$ and $\Delta_+ = \bar{\slashed{\mathcal D}}_{cl}\slashed{\mathcal D}_{cl}$. $\Delta_-$ has zero modes in self-dual configurations which are excluded from the determinants (denoted with a $\det^\prime$). In 3D, the contributions of nonzero modes do not cancel even in supersymmetric theories \cite{Dorey:1997ij,Dorey:1998kq}. As a result, the path integral of
the $\mathcal{N}=2$ theory with $F$ fundamental flavors is proportional to a
factor of $\left(R_{adj}\right)^{3/4}\left(R_{fund}\right)^{-F/2}$,
where the ratio of determinants $R_\mathcal{R}$ for an arbitrary
representation $\mathcal{R}$ is given by
\begin{equation}
    R_{\mathcal R} = \frac{\det \Delta_+}{\det^\prime \Delta_-} = \lim_{\mu\rightarrow 0} \left [\mu^{I_\mathcal{R}(0)} \exp\left(-\int_{\mu^2}^\infty \frac{dM^2}{M^2}I_{\mathcal{R}}(M^2)\right)\right ]\,.
\end{equation}
$I_{\mathcal R}(M^2)$ is the generalized zero mode index for representation $\mathcal R$. The ratio of nonzero mode determinants is $R_{adj}=(2v)^4$ for the adjoint and $R_{fund}=v^2$ for the fundamental representation \cite{Dorey:1998kq}.

After converting the zero mode integrals to collective coordinates and using zero mode solutions normalized to one, the path integral measure becomes \cite{Dorey:1997ij}
\begin{equation}
    \int [D\phi_0]= \int \frac{d^3z}{(2\pi)^{3/2}}(S_{cl})^{3/2}\int \frac{d\theta}{(2\pi)^{1/2}} \left(\frac{S_{cl}}{v^2} \right)^{1/2} \int d^2 \xi \int d\chi \int d\bar\chi\,,
\end{equation}
where $z$ and $\theta$ make up the bosonic collective coordinates, and the  $\xi, \chi$ and $\bar\chi$ are Grassmannian collective coordinates for the gauginos and quarks respectively.
If we expand the effective squark action in supersymmetric gauge couplings, the correlation function simplifies to
\begin{multline}
    \expval{\bar q^i(x) q_i(x)} = \frac{(R_{adj})^{3/4}(R_{fund})^{-1/2}}{2\pi} \frac{(S_{cl})^2}{v} e^{-S_{cl}} \int d^3z \int d^2\xi \int d\chi d\bar\chi \\
    \times \bar q(x) q(x) \int d^3 y_1 \left (q^* \lambda \psi \right )\int d^3 y_2 \left (\bar q^* \lambda \bar \psi \right )\,. \label{eq:simpint}
\end{multline}
Now we turn to deriving the fermionic zero modes.

\subsection{Gaugino Zero Modes}

Reverting to $\sigma^a$ being the fourth component of the four-vector gauge field $A_\mu$, the gaugino has zero mode solutions resulting from supersymmetry transformations on the monopole field configuration
\begin{equation}
    \lambda\indices{^a_\alpha^{[\beta]}}(r) = \frac{-1}{\sqrt{2}}(\sigma^{\mu\nu})\indices{_\alpha^\beta} F\indices{^a_{\mu\nu}} \equiv \sqrt{2} (\sigma^k)\indices{_\alpha^\beta}B_k^a(r)\,,
\end{equation}
where $B_k^a(r)$ is the $k$th component of the $a$th color magnetic field, $B_k^a = -\frac{1}{2}\epsilon_{ijk}F^{a\,ij}$, in monopole background and $\beta$ labels the two zero modes. Explicit evaluation finds
\begin{equation}
    B_k^a(r) = (\delta^a_k - n^k n^a ) \frac{v H(vr)}{r}(1-F(vr)) + n^k n^a \frac{F(vr)}{r^2}(2-F(vr))\,.
\end{equation}
After normalizing the gaugino zero modes and introducing a dimensionless function, $\tilde B_k^a(vr) = \frac{1}{v^2} B_k^a$, we find
\begin{equation}
    \lambda\indices{^a_\alpha^{[\beta]}}(r) = \sqrt{\frac{g^2v^3}{4\pi}} (\sigma^k)\indices{_\alpha^\beta}\tilde B_k^a(vr)\,. \label{eq:gaugino}
\end{equation}

\subsection{Quark Zero Modes}

Zero mode solutions for fundamental fermions were found in \cite{Jackiw:1975fn} and are given by
\begin{equation}
    \psi_{i\alpha}(r) = (\sigma^2)_{i\alpha} C \exp\bigg [-\int_0^r dr \bigg (\frac{v}{2}H(vr) + \frac{F(vr)}{r} \bigg ) \bigg ]\,,
\end{equation}
where $C$ is the normalization constant. In supersymmetric gauge theories, there is a closed form solution,
\begin{equation}
    \psi_{i\alpha}(r) = (\sigma^2)_{i\alpha}\sqrt{\frac{v^3}{8\pi}} \frac{\tanh \frac{vr}{2}}{\sqrt{vr\sinh vr}} := (\sigma^2)_{i\alpha}\sqrt{\frac{v^3}{8\pi}} X(vr)\,, \label{eq:quark}
\end{equation}
where $X(r)$ is implicitly defined. Similar solutions for the antifundamental modes can be found by raising indices with the antisymmetric tensor, $\epsilon^{ij}$. Reintroducing the Grassmannian coordinate, $\chi$, to the fermion field, the zero mode is $(\psi_0)_{i\alpha} = \psi_{i\alpha}\chi$.

\subsection{Evalulating the integral}

Inserting \eqref{eq:gaugino} and \eqref{eq:quark} into \eqref{eq:simpint} then performing the Grassmann integration and replacing the products of squark operators with their Green's functions,\footnote{The integral should be dominated by regions where the squarks are essentially free, massive fields. In these regions, the Green's function is $q^*_i(x)q^j(y)=G_i^j(\abs{x-y})\sim\delta_i^j e^{-v\abs{x-y}/2}(4\pi\abs{x-y})^{-1}$.} one finds
\begin{equation}
    \expval{\bar q^i(x) q_i(x)} = \frac{8v^2}{2\pi} \frac{(4\pi)^2 v}{g^4} \frac{g^2 v^8}{32\pi^2} e^{-S_{cl}} \int d^3z \prod_{i=1,2} \int \frac{d^3 y_i}{4\pi} \frac{e^{-\frac{v}{2}\abs{x-y_i}}}{v\abs{x-y_i }} \Omega(v\abs{z-y_i }) \,,
\end{equation}
where $\Omega(\rho)=\delta^k_aX(\rho)\tilde B^a_k(\rho)$. Shifting the center of integration such that $y_i \rightarrow y_i + z$ and $z\rightarrow z+x$, the $x$ dependence drops out.
The angular $y_i$ integrals can be evaluated and the integral simplifies to
\begin{equation} \label{eq:nonholans}
    \ev{\bar q^i(x) q_i(x)} = \frac{v^2}{g^4} e^{-S_{cl}} g^2 I\,,
\end{equation}
where $I = 4\int d\rho_z \prod_{i=1,2} \int d\rho_i \rho_i (e^{-\abs{\rho_z-\rho_i}/2}-e^{-\abs{\rho_z+\rho_i}/2}) \Omega(\rho_i)$. Note that the $\rho_i$ are the dimensionless magnitudes of the 3D vectors, $\rho_i=v\abs{\vec{y_i}}$. It takes some work, but one can show that the integrand of $I$ is positive definite and converges quickly. Thus our answer is
\begin{equation}
    \ev{\bar q^i(x) q_i(x)} \sim g^2 \left(\frac{v^2}{g^4}\right) e^{-S_{cl}} \,,
\end{equation}
which is nonholomorphic due to the factor of $v^2/g^4$. As explained in  \cite{Dorey:1998kq,Poppitz:2012sw}, this nonholomorphic factor reflects finite renormalization of $g^2$ and can be absorbed into the definition of the kinetic terms in the low energy theory. After taking this into account, the correlation function becomes a holomorphic relation between chiral operators $YM=g^2$.

\end{document}